\documentclass[12pt,preprint]{aastex}

\def\kms{km~s$^{-1}$}

\def\msun{$M_{\odot}$}
\def\rsun{$R_{\odot}$}

\def\hbeta{H$\beta$}

\def\Te{T_{\rm eff}}
\def\logg{\log g}
\def\gta{\lower 0.5ex\hbox{$\buildrel > \over \sim\ $}} 
\def\lta{\lower 0.5ex\hbox{$\buildrel < \over \sim\ $}} 

\slugcomment{\bf Accepted for Ap.J.}

\begin{document}

\title{Analysis of a Very Massive DA White Dwarf via the Trigonometric Parallax
and Spectroscopic Methods}

\author{Conard C. Dahn }
\affil{US Naval Observatory, Flagstaff Station, P.O. Box 1149, Flagstaff, AZ 86002-1149}
\email{dahn@nofs.navy.mil}

\author{P. Bergeron }
\affil{D\'epartement de Physique, Universit\'e de Montr\'eal,
C.P. 6128, Succ. Centre-Ville, Montr\'eal, Qu\'ebec,
Canada H3C 3J7 }
\email{bergeron@astro.umontreal.ca}

\author{James Liebert }
\affil{Steward Observatory, University of Arizona, Tucson AZ 85726} 
\email{liebert@as.arizona.edu}  

\author{Hugh C. Harris}
\affil{US Naval Observatory, Flagstaff Station, P.O. Box 1149, Flagstaff, AZ 86002-1149}
\email{hch@nofs.navy.mil}

\author{and}

\author{S. K. Leggett}
\affil{UKIRT, Joint Astronomy Centre, 660 North A'ohoku Place, Hilo, HI 96720}
\email{s.leggett@jach.hawaii.edu}

\begin{abstract}

By two different methods, we show that LHS 4033 is an extremely
massive white dwarf near its likely upper mass limit for destruction
by unstable electron captures. From the accurate trigonometric
parallax reported herein, the effective temperature ($\Te=10,900$~K)
and the stellar radius ($R=0.00368$ \rsun) are directly determined
from the broad-band spectral energy distribution --- the parallax
method.  The effective temperature and surface gravity are also
estimated independently from the simultaneous fitting of the observed
Balmer line profiles with those predicted from pure-hydrogen model
atmospheres --- the spectroscopic method ($\Te=10,760$~K,
$\logg=9.46$). The mass of LHS 4033 is then inferred from theoretical
mass-radius relations appropriate for white dwarfs.  The parallax
method yields a mass estimate of 1.310--1.330
\msun, for interior compositions ranging from pure magnesium to pure
carbon, respectively, while the spectroscopic method yields an estimate of
1.318--1.335 \msun\ for the same core compositions.  This star is the
most massive white dwarf for which a robust comparison of the two
techniques has been made.

\end{abstract}


\section{INTRODUCTION}

LHS 4033 (WD 2349$-$031) is a white dwarf discussed in a recent paper
by \citet{salim04}. The star has also been part of the Luyten Half
Second (LHS) survey $\mu \ge 0.6^{\prime\prime}$ yr$^{-1}$ white dwarf
sample of C. C. Dahn, H. C. Harris, S. K. Leggett, \& J. Liebert
(2004, in preparation), virtually all of which have been targeted for
accurate trigonometric parallaxes at the U.S. Naval Observatory, for
purposes of estimating the luminosity function of cool white dwarfs.

In the H-R diagram, this object lies to the left of the diagonal
sequence of white dwarfs, indicating that its mass is larger and
its radius is smaller than normal.  In this paper we show that its
mass is, indeed, extraordinarily large for a white dwarf.
A few other stars of mass similar to the value we
present in this paper for LHS 4033 have been found -- i.e. the
low-field magnetic star PG 1658+441 \citep[][$M \sim 1.31$
\msun]{schmidt92,dupuis03}, and the highly-magnetic white dwarf
RE J0317$-$853 \citep[][$M=1.35$ \msun]{barstow95,ferrario97}.
Note that the application of the Balmer line fitting procedure to
these two magnetic stars of high mass is impossible for a $\sim$300~MG
magnetic white dwarf like RE J0317$-$853, and the mass of this star
has been determined indirectly from the companion. Also, the modelling
of the Zeeman triplet components of PG 1658+441 by \citet{schmidt92}
is only approximate, and it is inherently less accurate due, among
other things, to the lack of a rigorous theory of Stark broadening in
the presence of the 3.5~MG field. Hence, both mass estimates are quite
uncertain.

Since the spectral type of LHS 4033 is DA and non-magnetic, the mass
may be estimated by fits to the Balmer lines \citep[see, e.g.,][]{BSL}
in a much more rigorous fashion. The surface gravity used with
suitable evolutionary models yields independent determinations of the
mass and radius. The effective temperature may also be estimated from
broad-band photometry once the dominant atmospheric constituent is
known.  This, along with an accurate trigonometric parallax, permits a
different estimate of the luminosity, radius, and mass
\citep{BLR}. While it has been possible to compare the parameter
determinations of these methods for limited samples of white dwarfs,
it is particularly interesting to do so for a massive star.

\section{OBSERVATIONAL DATA}

Photometry with $BVI$ filters was obtained three times with the USNO
1.0 m telescope generally during 1997-1998. $JHK$ data were obtained
on 1998 October 12 using the IRCAM camera outfitted with the UKIRT
system filters and calibrated using UKIRT standards
\citep{hawarden01}.  Colors are reduced to the Johnson system for
$B$--$V$, the Cousins system for $V$--$I$, and the CIT system for
$J$--$H$ and $H$--$K$. Errors are 0.02 mag in $BVI$ and 0.05 in
$JHK$. Our optical and infrared photometry for LHS 4033 is given in
Table 1. \citet{salim04} also report CCD photometry for LHS
4033, on the Johnson-Cousins system, from the Lick Observatory 1 m
Nickel telescope. They obtain $B=17.162\pm0.020$, $V=16.992\pm0.017$,
$R=16.987\pm0.030$, and $I=16.936\pm0.036$, based on 2-3 observations
per band. The corresponding color indices, $B$--$V=0.17$ and
$V$--$I=0.056$, thus agree with our measurements within the
uncertainties. In the model atmosphere analysis presented below, we
rely on our own photometric measurements only.

Trigonometric parallax observations were carried out over a 6.05 year
interval (1997.76 -- 2003.81) using the USNO 1.55 m Strand Astrometric
Reflector equipped with a Tek2K CCD camera (Dahn 1997).  The absolute
trigonometric parallax and the relative proper motion and position
angle derived from the 150 acceptable frames are given in Table 1.
The parallax and apparent V magnitude then yield an absolute
magnitude, also included in Table 1.  Further details regarding the
astrometry for LHS 4033 will appear in a paper on white dwarf
parallaxes (Dahn et al. 2004, in preparation).

Finally, optical spectroscopy was secured on 2003 October 1 using the
Steward Observatory 2.3-m reflector telescope equipped with the Boller
\& Chivens spectrograph and a UV-flooded Texas Instrument CCD
detector. The 4.5 arcsec slit together with the 600 lines mm$^{-1}$
grating blazed at 3568 \AA\ in first order provided a spectral
coverage of 3120--5330 \AA\ at an intermediate resolution of $\sim
6$~\AA\ FWHM. The 3000 s integration yielded a signal-to-noise ratio
around 55 in the continuum. Our optical spectrum for LHS 4033 is
contrasted in Figure
\ref{fg:f1} with that of G61$-$17, a DA white dwarf with an effective
temperature comparable to that of LHS 4033, but with a normal surface
gravity and mass ($\Te=10,680$~K, $\logg=8.06$, $M=0.64$ \msun)
according to the spectroscopic analysis of the DA stars from the PG
sample by J. Liebert, P., Bergeron, \& J. B. Holberg (2003, in
preparation). The strong decrement of the high Balmer lines already
indicates that LHS 4033 is a massive white dwarf.

\section{MODEL ATMOSPHERE ANALYSIS}

\subsection{Photometric Analysis}

We first proceed to fit the optical and infrared photometry using the
technique described in Bergeron et al.~(1997, 2001). Broadband
magnitudes are first converted into observed fluxes using Eq.~[1] of
\citet{BRL} with the appropriate zero points.  The resulting energy
distribution is then compared with those predicted from our model
atmosphere calculations, with the monochromatic fluxes properly
averaged over the same filter bandpasses. 

Our model atmospheres are hydrogen-line blanketed LTE models, and
assume a pure hydrogen composition.  Convection is treated within the
mixing-length theory, with the ML2/$\alpha=0.6$ formulation following
the prescription of \citet{bergeron95}. The calculations of
theoretical spectra are described at length in \citet{bergeron91b},
and include the occupation probability formalism of \citet{HM88}. This
formalism allows a detailed calculation of the level populations in
the presence of perturbations from neighboring particles, and also
provides a consistent description of bound-bound and bound-free
transitions.

The observed fluxes $f_{\lambda}^m$ and Eddington model fluxes
$H_{\lambda}^m$ --- which depend on $\Te$ and log $g$ --- for a given
bandpass $m$ are related by the equation

$$f_{\lambda}^m= 4\pi~(R/D)^2~H_{\lambda}^m\ , \eqno (1)$$

\noindent where $R/D$ is the ratio of the radius of the star to its distance
from Earth.  Our fitting technique relies on the nonlinear
least-squares method of Levenberg-Marquardt \citep{press86}, which is
based on a steepest descent method.  The value of $\chi ^2$ is taken
as the sum over all bandpasses of the difference between both sides of
Eq.~[1], properly weighted by the corresponding observational
uncertainties. Only $\Te$ and the solid angle $\pi~(R/D)^2$ are
considered free parameters, while the uncertainties 
are obtained directly from the covariance matrix of the fit. 

We first assume $\logg=8.0$ and determine the effective temperature
and the solid angle, which, combined with the distance $D$ obtained
from the trigonometric parallax measurement, yields directly the
radius of the star $R$. The latter is then converted into mass using
an appropriate mass-radius relation for white dwarf stars. Here we
first make use of the mass-radius relation of \citet{hamada} for
carbon-core configurations.  This relation is preferred to the
evolutionary models of \citet{wood95} or those of \citet{fon01}, which
extend only up to 1.2 and 1.3 \msun, respectively. Uncertainties due
to finite temperature effects and core composition will be discussed
below. In general, the value of $\logg$ obtained from the inferred
mass and radius ($g=GM/R^2$) will be different from our initial
assumption of $\logg=8.0$, and the fitting procedure is thus repeated
until an internal consistency in $\logg$ is achieved. The parameter
uncertainties are obtained by propagating the error of the photometric
and trigonometric parallax measurements into the fitting procedure.

Our best fit to the optical $BVI$ and infrared $JHK$ photometry of LHS
4033 is displayed Figure \ref{fg:f2}. The monochromatic fluxes from
the best fitting model are shown here as well, although the formal fit
is performed using only the average fluxes (filled dots). The solution
at $\Te=10,900\pm290$~K and $R=0.00368\pm0.00013$ \rsun\ implies a
stellar mass of $M=1.330\pm0.004$ \msun\ and a value of
$\logg=9.43\pm0.02$. The parameters of both methods are summarized in
Table 2. The predicted absolute visual magnitude obtained from the
values of $\Te$ and $\logg$ is $M_V=14.63$, in perfect agreement with
the value derived from the parallax given in Table 1.

\subsection{Spectroscopic Analysis}

The optical spectrum of LHS 4033 is fitted with the same grid of
model atmospheres following the procedure described in
\citet{BSL} and \citet{bergeron95}. The spectrum is first fitted with
several pseudo-Gaussian profiles \citep{saffer88} using the nonlinear
least-squares method of Levenberg-Marquardt described above.  Normal
points defined by this smooth function are then used to normalize the
line flux to a continuum set to unity at a fixed distance from the
line center. The comparison with model spectra, which are convolved
with the appropriate Gaussian 6 \AA\ instrumental profile, is then
carried out in terms of these normalized line profiles only. Our minimization
technique again relies on the Levenberg-Marquardt method using the
\hbeta\ to H8 line profiles. Our best fit is displayed in Figure
\ref{fg:f3}.

Remarkably, our spectroscopic solution $\Te=10,760\pm150$~K and
$\logg=9.46\pm0.04$, which translates into $M=1.335\pm0.011$ and
$R=0.00358\pm0.00019$ \rsun\ using the Hamada-Salpeter mass-radius
relation for carbon-core configurations, is in excellent agreement with
the solution obtained with the photometry and trigonometric parallax
method. This is arguably the most massive white dwarf subjected to a
rigorous mass determination \citep[see, e.g., Table 3
of][]{dupuis02}. Note that despite the extreme surface gravity of LHS
4033, the Hummer-Mihalas formalism used in the line profile calculations
remains perfectly valid, since the density at the photosphere remains low
($\rho\sim10^{-5}$ g cm$^{-3}$) as a result of the high opacity of
hydrogen at these temperatures.

\subsection{Mass-Radius Relation}

In a venerable paper, \citet{hamada} first employed an
equation-of-state (EOS) including coulomb ``corrections'' to the
pressure and energy of a degenerate Fermi gas \citep{salpeter61} to
calculate the mass-radius-central density relations for models
composed of helium through iron. These corrections to the classic
Chandrasekhar EOS for degenerate matter are more important at high
mass. It may also be noted that, especially at the relatively low
effective temperature of LHS 4033, neglect of the internal energy of
the ions (``zero-temperature'' modelling) is likely to be a reasonable
assumption. 

Since LHS 4033 may have a core composed of material much heavier than
carbon, we must explore the effects of core composition on the results
of our analysis. We compare in Figure \ref{fg:f4} the mass-radius
relation obtained from the detailed evolutionary carbon- and
carbon/oxygen-core models of \citet[][see also Bergeron et
al. 2001]{fon01} with the Hamada-Salpeter zero-temperature
configurations for carbon and magnesium at a mass of 1.3 \msun, the
highest mass of the Fontaine et al.~models. At the effective
temperature and mass of LHS 4033, the carbon- or
carbon/oxygen-core models of Fontaine et al.~reveal that finite
temperature effects are extremely small, and account for an increase
in radius of only $\sim 0.5$ \% (i.e.~by comparing the radius at
10,000 K with the value at 3500 K where it becomes constant).
Moreover, at the temperature of LHS 4033, the carbon-core models of
Fontaine et al.~and Hamada-Salpeter differ by only 2.7 \% in radius,
or 0.007 \msun\ in mass. Details of the equation-of-state are thus
also negligible in the present context. Finally, the Mg and C
configurations of Hamada-Salpeter differ by 7.4 \% in radius, or 0.02
\msun\ in mass. Indeed, the parallax method with the Mg configurations
yields a mass of 1.310
\msun\ (instead of 1.330 when C configurations are used),
while the spectroscopic method yields a mass of 1.318 \msun\ (instead
of 1.335 \msun). These results are also reported in Table 2. We thus
argue that our mass estimates are uncertain by 0.02 \msun\ at the
most.

\section{DISCUSSION}

The theoretical prediction is that, with a mass near 1.33 \msun, LHS
4033 has been through carbon burning, {\it if} it has evolved as a
single star.  Perhaps more plausible is the possibility that massive
white dwarfs apart from those found in young clusters or associations
are generally the results of mergers of more ordinary, probably C-O
white dwarfs \citep[cf.][]{bergeron91a,vennes96,marsh97}. In the
former case, LHS 4033 can be expected to have an interior composed of
some mix of O$^{16}$, Ne$^{20}$, Mg$^{24}$ and even Na$^{23}$
\citep[an O-Ne-Mg core;][and references therein]{garcia94,ritossa99}.
For our mass determination of $1.33-1.34$ \msun, the Hamada-Salpeter
calculations place LHS 4033 very near their predicted maximum mass of
$1.363$ \msun\ for Mg$^{24}$ and $1.396$ \msun\ for C (with other
species with atomic weights intermediate between these extremes having
values presumably in between).  Thus, there is little difference
between the assumptions of an O-Ne-Mg interior (an unusual single
star), and that of a normal C-O composition (binary origin).  Above
this mass limit -- significantly less than that of Chandrasekhar near
1.4 \msun\ -- electron captures (traditionally called inverse beta
decays) will produce increasingly neutron-rich nuclei, increasing the
mean molecular weight per particle, and resulting in some kind of
detonation of the core \citep[see, e.g.,][]{arnett96}.

The high mass and small radius cause the gravitational redshift of LHS
4033 to be much larger than for a normal white dwarf: $v_{\rm GR}=
0.635\,(M/M_{\odot})\,(R_{\odot}/R) = 237$ \kms.  This fact accounts for
the high radial velocity (206 \kms) observed by \citet{salim04}
without requiring that the star have a high space velocity.  The
tangential velocity, using the proper motion and distance reported in
Table 1, is 97 \kms, and also is not as large as would be derived from
a photometric distance estimate assuming the star had a normal gravity.
The components of its space velocity, based on the proper motion in
Table 1 and the radial velocity observed by Salim et al. corrected for
the gravitational redshift above, are $(U,V,W) = (87, -11, 31)$ \kms.
(These values have been corrected for the Sun's peculiar velocity,
and $U$ is away from the Galactic center and V is in the direction
of Galactic rotation.)  This space velocity is consistent with the
kinematics of the Galaxy's old disk.

\acknowledgments {We thank S.~Boudreault for the reduction of the optical 
spectrum, and G. Fontaine for useful discussions.
The astrometric observations were carried out as part of the USNO
parallax program, and we thank the following individuals who
contributed some of the frames used here --
Blaise Canzian, Harry Guetter, Stephen Levine, Chris Luginbuhl,
Alice Monet, Dave Monet, Ron Stone, and Dick Walker.  This work was
supported in part by the NSERC Canada and by the Fund FQRNT
(Qu\'ebec)}. 
JL acknowledges support from the National Science Foundation 
through grant AST-0307321 for study of white dwarfs.
UKIRT, the United Kingdom Infrared Telescope, is operated
by the Joint Astronomy Centre on behalf of the U.K. Particle Physics
and Astronomy Research Council.

\clearpage
\begin{deluxetable}{lc}
\tabletypesize{\footnotesize}
\tablecolumns{2}
\tablewidth{0pt}
\tablecaption{Photometric and Astrometric Data for LHS 4033}
\tablehead{
\colhead{Parameter} &
\colhead{Value}} 
\startdata
$V$      \dotfill & $16.98\pm0.02$ \\
$B$--$V$ \dotfill & $+0.19\pm0.03$ \\
$V$--$I$ \dotfill & $+0.07\pm0.03$ \\
$J$      \dotfill & $16.97\pm0.05$ \\
$J$--$H$ \dotfill & $+0.05\pm0.07$ \\
$H$--$K$ \dotfill & $-0.10\pm0.07$ \\
$\pi_{\rm abs}$ (mas) \dotfill & $33.9\pm0.6$ \\
$\mu_{\rm rel}$ (mas yr$^{-1}$) \dotfill & $701.4\pm0.2$ \\
PA (deg) \dotfill & $66.3\pm0.1$ \\
Distance (pc) \dotfill & $29.5\pm0.5$ \\
$M_V$                 \dotfill & $14.63\pm0.04$ \\
\enddata
\end{deluxetable}

\clearpage
\begin{deluxetable}{clllll}
\tabletypesize{\footnotesize}
\tablecolumns{6}
\tablewidth{0pt}
\tablecaption{Atmospheric Parameters of LHS 4033}
\tablehead{
\colhead{Core} &
\colhead{Method} &
\colhead{$\Te$ (K)} &
\colhead{$\logg$} &
\colhead{$R/$\rsun} &
\colhead{$M/$\msun}}
\startdata
C & Parallax & $10,900\pm290$ & $9.43\pm0.02$ & $0.00368\pm0.00013$ &
$1.330\pm0.004$ \\
& Spectroscopy & $10,760\pm150$ & $9.46\pm0.04$ & $0.00358\pm0.00019$ &
$1.335\pm0.011$ \\
\\
Mg & Parallax & $10,900\pm290$ & $9.42\pm0.02$ & $0.00368\pm0.00013$ &
$1.310\pm0.004$ \\
 & Spectroscopy & $10,760\pm150$ & $9.46\pm0.04$ & $0.00355\pm0.00019$ &
$1.318\pm0.011$ \\
\enddata
\end{deluxetable}

\clearpage

\figcaption[f1.eps] {Comparison of the optical spectrum of 
LHS 4033 with that of G61$-$17, a white dwarf with comparable effective
temperature ($\Te=10,680$~K) but with a normal mass of $M=0.64$ \msun\
($\logg=8.06$).  The spectra are normalized at 4400 \AA\ and are
shifted vertically by 0.5 for clarity.\label{fg:f1}}

\figcaption[f2.eps] {Fits to the energy distribution of LHS 4033
with pure hydrogen models. The optical $BVI$ and infrared $JHK$
photometric observations are shown by the error bars. The solid line
corresponds to the model monochromatic fluxes, while the filled
circles represent the average over the filter bandpasses (including
the model prediction at $R$).\label{fg:f2}}

\figcaption[f3.eps] {Model fits to the individual Balmer line profiles
of LHS 4033. All lines are normalized to a continuum set to unity and
offset vertically from each other by a factor of 0.2. Values of $\Te$
and $\logg$ have been determined from ML2$/\alpha=0.6$ models, while
the stellar mass has been derived from the mass-radius relation of
\citet{hamada} for carbon-core configurations.\label{fg:f3}}

\figcaption[f4.eps] {Comparison of the mass-radius relation at $M=1.3$ 
\msun\ obtained from the evolutionary carbon- and carbon/oxygen-core models of 
\citet[][FBB/C and FBB/C-O, respectively]{fon01} and the carbon and magnesium
zero-temperature configurations of \citet[][HS/C and
HS/Mg, respectively]{hamada}.\label{fg:f4}}

\clearpage
\begin{figure}[p]
\plotone{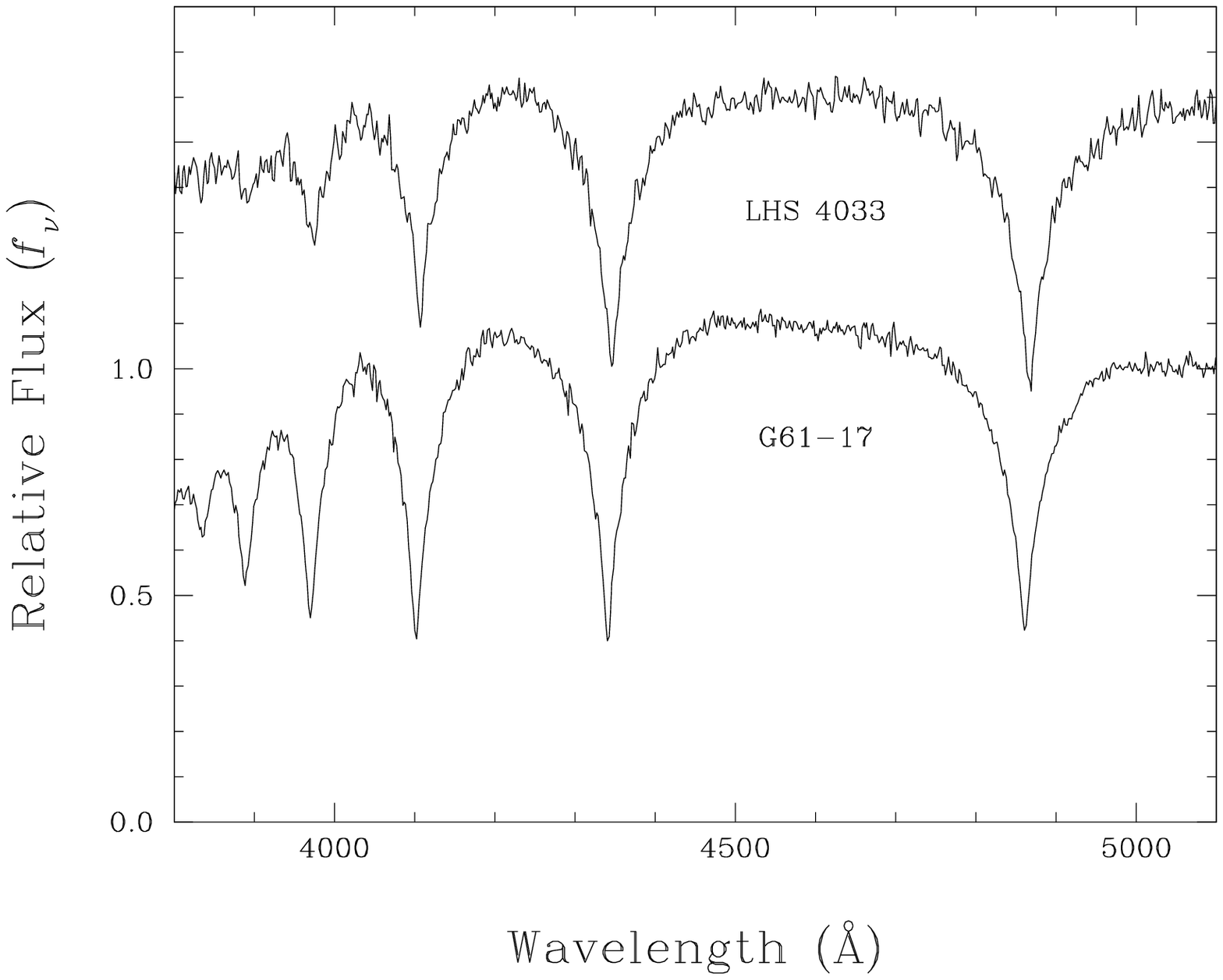}
\begin{flushright}
Figure \ref{fg:f1}
\end{flushright}
\end{figure}

\clearpage
\begin{figure}[p]
\plotone{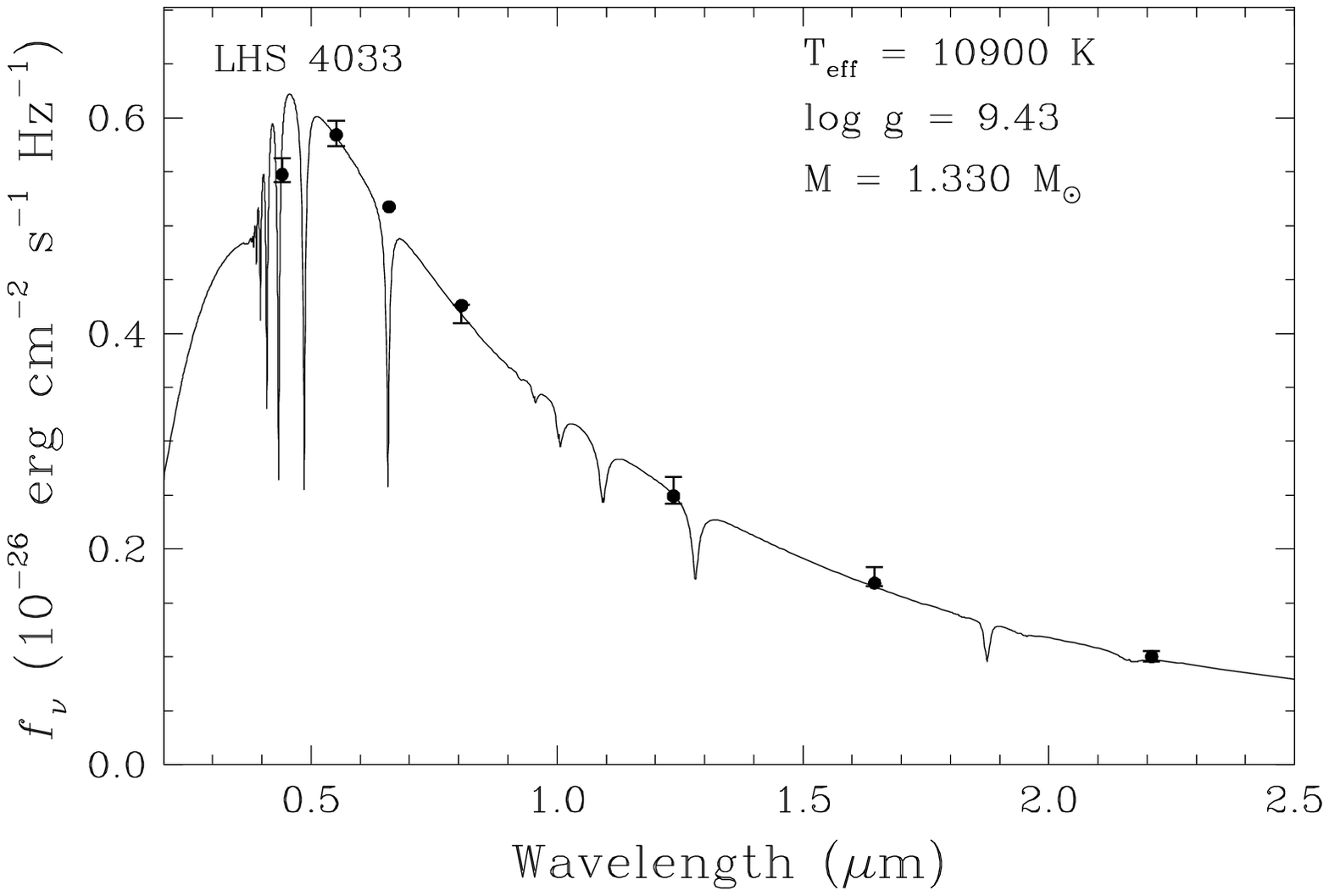}
\begin{flushright}
Figure \ref{fg:f2}
\end{flushright}
\end{figure}

\clearpage
\begin{figure}[p]
\plotone{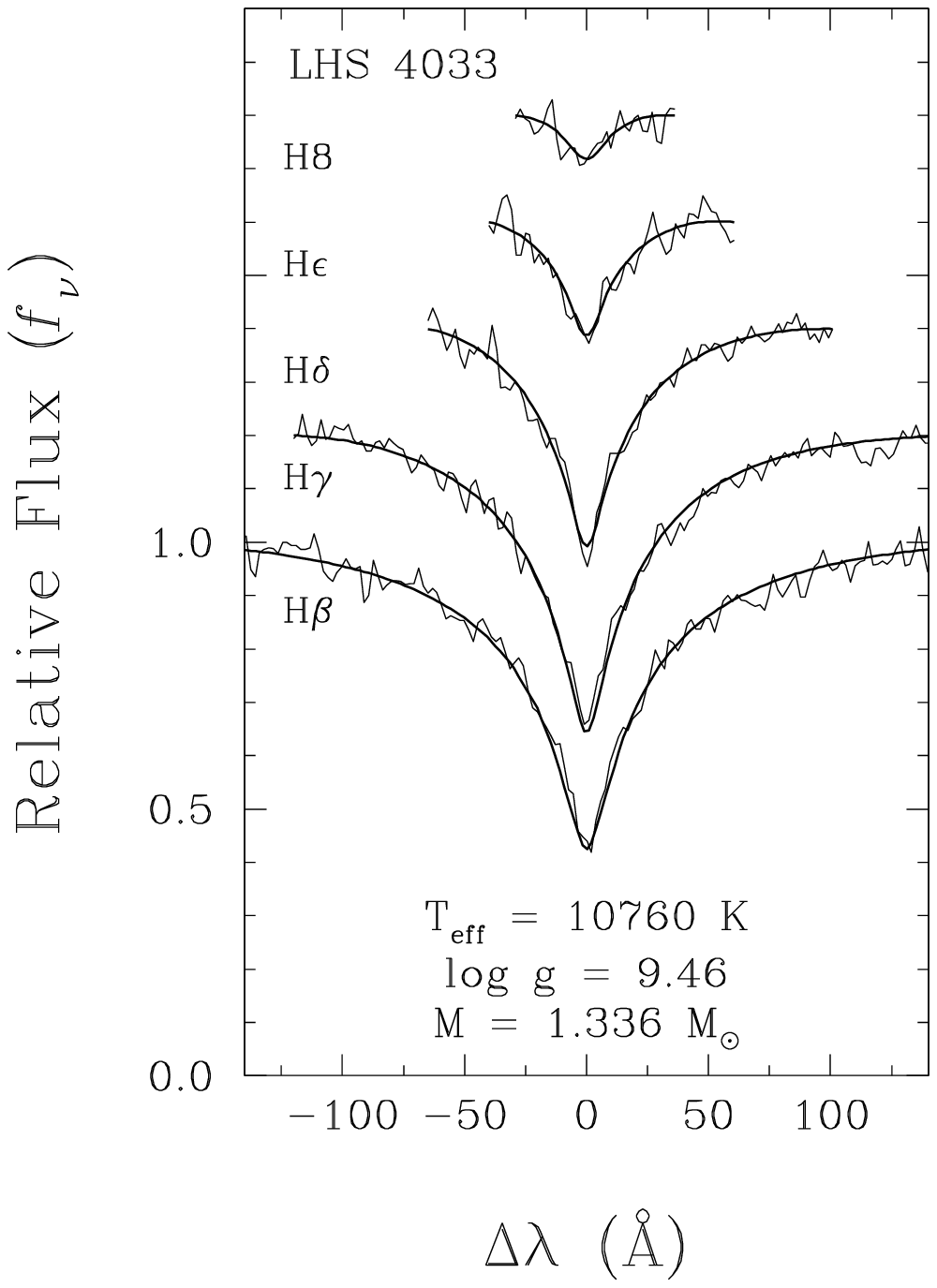}
\begin{flushright}
Figure \ref{fg:f3}
\end{flushright}
\end{figure}

\clearpage
\begin{figure}[p]
\plotone{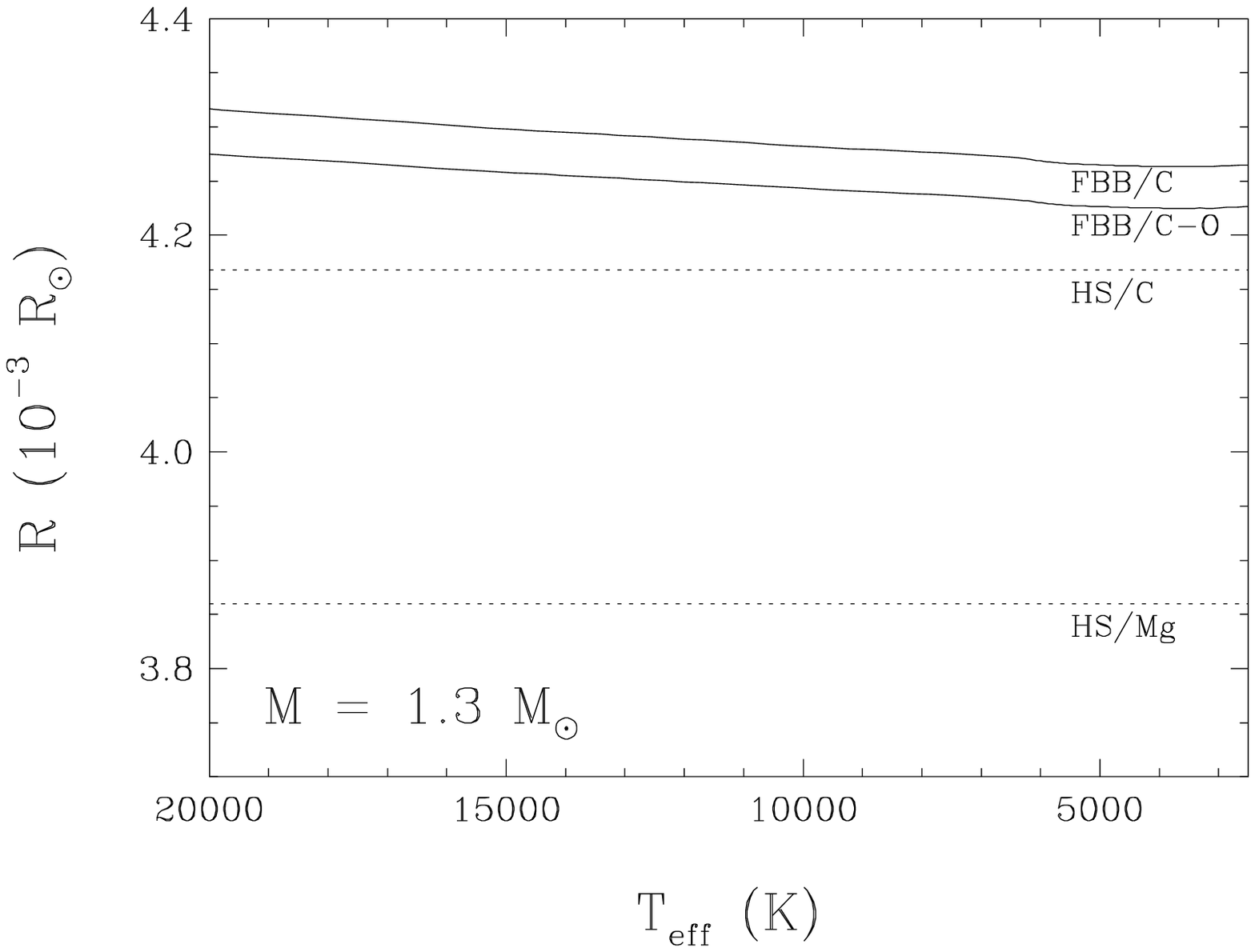}
\begin{flushright}
Figure \ref{fg:f4}
\end{flushright}
\end{figure}

\end{document}